%% file: main.tex
\documentclass[fleqn,10pt]{wlscirep}
\usepackage[utf8]{inputenc}
\usepackage[T1]{fontenc}
\usepackage{lineno}
\usepackage{float}

\title{A Comprehensive Indoor Environment Dataset from Single-family Houses in the US}

\author[1]{Sheik Murad Hassan Anik}
\author[2]{Xinghua Gao}
\author[1]{Na Meng}
\affil[1]{Department of Computer Science, Virginia Tech, Blacksburg, VA 24061, USA}
\affil[2]{Myers-Lawson School of Construction, Virginia Tech, Blacksburg, VA 24061, USA}

\affil[*]{corresponding author(s): Xinghua Gao (xinghua@vt.edu)}

\input{contents/1_abstract}

\begin{document}

\flushbottom
\maketitle

\thispagestyle{empty}

\input{contents/2_background}
\input{contents/3_methods}
\input{contents/4_data_records}

\input{contents/5_0_data_distribution}
\input{contents/5_technical_validation}
\input{contents/6_usage_notes}
\input{contents/7_code_availability}
\bibliography{references}

\section*{Author contributions statement}
\input{contents/8_author_contributions}

\section*{Competing interests} 
\input{contents/9_competing_interests}

\end{document}

%% file: contents/1_abstract.tex
\begin{abstract}
The paper describes a dataset comprising indoor environmental factors such as temperature, humidity, air quality, and noise levels. The data was collected from $10$ sensing devices installed in various locations within three single-family houses in Virginia, USA. The objective of the data collection was to study the indoor environmental conditions of the houses over time. The data were collected at a frequency of one record per minute for a year, combining over $2.5$ million records. The paper provides actual floor plans with sensor placements to aid researchers and practitioners in creating reliable building performance models. The techniques used to collect and verify the data are also explained in the paper. The resulting dataset can be employed to enhance models for building energy consumption, occupant behavior, predictive maintenance, and other relevant purposes.
\end{abstract}

%% file: contents/2_background.tex
\section{Background \& Summary}
Data generated in the built environment can provide a comprehensive understanding of various aspects of a building, enabling informed decision-making, improving occupant experiences, and enhancing overall building performance and sustainability. Data related to energy consumption, Heating, Ventilation, and Air Conditioning (HVAC) systems, lighting, and water usage can help identify inefficiencies and opportunities for optimization. This information allows for more informed decision-making regarding system upgrades, retrofits, or maintenance schedules. Data on how occupants interact with a building's spaces, systems, and technologies can reveal patterns, preferences, and habits. This information can be used to improve occupant comfort, well-being, and productivity by adjusting environmental factors, such as temperature, lighting, and air quality, according to their needs and preferences. 
\\

The accessibility of open datasets is of paramount importance in the investigation of indoor residential environments and their respective habitats. Such availability expedites research on housing performance and fosters the development of advanced energy analysis methodologies. The built environment not only exerts active influences but also yields passive effects on human productivity. It bears a significant impact on the health and comfort levels of the inhabitants within the residential setting \cite{hedge2014occupant, mirzaei2020impact, andargie2019review, zhang2019healing, ghodrati2012green, mujan2019influence} as individuals spend 87\% of their time in indoor environments \cite{mujan2019influence}. Consequently, it is imperative to maintain optimal conditions to enhance the overall experience of both the built environment and its inhabitants. 
\\

Open indoor environmental data possesses considerable potential to enhance architectural designs, simulations, and, inform future decisions pertaining to building construction, operations management, and innovative design approaches. It facilitates the establishment of performance evaluation benchmarks across diverse geographic locations, building ages, and typologies. \cite{li2014methods, roth2020examining}. These datasets can further contribute to the benchmarking of machine learning models in the context of building and habitat data analysis, thereby promoting the development and evaluation of more accurate and robust predictive algorithms. \cite{quevedo2023applying, robinson2017machine}. Consequently, indoor environmental datasets have emerged as a critical component with the potential to enhance the overall built environment by providing novel and innovative development guidelines. These improvements encompass reducing building operating expenses and elevating the living experience within indoor environments. In recent years, the trend of developing open-source Building Performance Data (BPD) and the public dissemination of such datasets have garnered increasing interest and momentum \cite{pipattanasomporn2020cu, tasgaonkar2022indoor, yoon2022datasets, gao2022understanding, thorve2023high, schwee2019room, dong2022global, agee2021measured, paige2019fleece}. Advancements in the Internet of Things (IoT) technologies \cite{bashir2016towards, baghalzadeh2022internet, tang2019review} have brought forth new techniques and methods for leveraging indoor environmental datasets. Air quality sensing and monitoring systems proposed in Zakaria \textit{et al}. \cite{zakaria2018wireless} and Marques \textit{et al}. \cite{marques2019cost} can be useful for the habitats from multiple dimensions like detection of harmful gases, measuring optimal oxygen levels, creating alerts for the extensive presence of carbon monoxide, etc.
\\

The present study introduces a dataset comprising one year of indoor environmental data from three single-family houses in Virginia, USA. This dataset results from a unique case study, which was an integral component of a longitudinal, multiphase investigation designed to explore the various attributes of occupants and their environments. All participating households gave their informed consent to partake in the study, and the research was conducted in compliance with all Institutional Review Board (IRB) protocols. The indoor environmental data was collected from $10$ sensing devices placed in different households through the Building Data Lite (BDL) system \cite{anik2022cost}. The devices are composed of different sensors to accumulate different environmental information. They were deployed in these households around the summer of 2021 and collected data for about a year with the frequency of $1$ record every minute. The subsequent sections of this paper delineate the data collection procedures, the resulting data records, and, the technical validation of the collected data.

%% file: contents/3_methods.tex
\section{Methods}
This section elaborates on the different components of the data collection methods used in this study. Beginning with the data collection architecture, sensors used in the process, sensing device deployment, and, participant declaration. The study utilized the Building Data Lite (BDL) \cite{anik2022cost} system to collect data. A unique set of sensors were used to capture the different information from the surroundings. The sensing devices were deployed to multiple individual houses with the informed consent of the occupants. 

\subsection{Building Data Lite}
The data presented in this paper were obtained using the Building Data Lite (BDL) system \cite{anik2022cost}. BDL is a distributed, portable, scalable, and cost-effective indoor environment sensing system. The BDL system is an open-source platform \cite{AnikGitHub2021} that facilitates the development of customizable and portable sensing devices capable of connecting to a central server. These sensing devices continuously transmit collected data to the central server via the internet. However, in case of a connection interruption, the data is stored on the local storage available in each sensing device. The BDL system features a web interface through the central server to access and download the collected data. All data can be publicly accessed at \href{https://www.building-data-lite.com}{www.building-data-lite.com}. 
\\

The architecture of BDL consists of three parts, the sensing devices, the data transmission medium, and, the central server. The system is capable of integrating a practically indefinite number of sensing nodes. The sensing nodes are based on Raspberry Pi devices (RPi). Each sensing device (RPi) can integrate multiple sensors for accumulating different environmental information. The sensors connected to each RPi record instantaneous information from the surroundings and transmit the data to the connected RPi. Collected data is stored in the local database of the RPi device. The sensing devices connect to the central server using the Internet as a transmission medium. The latest RPi devices include both wireless modems and Ethernet ports that can be used for the connection. The sensing devices transmit the newly collected data to the central server every hour. If there consists any issue during the transmission like power-cut, no connectivity, or, server problem, the data remains in the local storage of the sensing device. In the next hour, the RPi tries to resend the data package. The frequency of transmission can also be modified according to needs. The central server is designed using PHP. It can be hosted both locally and on a live server. Depending on the setup of the central server, its URL needs to be changed in each sensing device for it to properly communicate with the server. The central server includes a web-based graphical user interface for no-code installation of sensing devices, visualization of collected data, and, data download. 
\\

The data collected in the BDL system is first stored in the local database of the sensing devices and then transmitted to the central database located in the central server. Both of these databases are relational databases and share similar characteristics. The RPi devices use the inbuilt MariaDB \cite{MariaDBFoundation2016, kenler2015mariadb} to store offline data. The central server utilizes MySQL \cite{bartholomew2012mariadb} to organize all collected data and share it through the visualization interface. 

\subsection{Sensor array}
This this phase of the data collection, the BDL system was utilized to collect data. The sensing nodes consisted of either an Enviro or an Enviro Plus sensor array \cite{gas_datasheet} along with Raspberry Pi Zero. Both of these sensor boards function similarly and include the same sensors except for the analog gas sensor which is exclusive to the Plus edition of Enviro. The following are the descriptions of the different sensors used in this study: 

\subsubsection{Temperature, pressure, and, humidity measurement}
The BME280 sensor \cite{sensortec2015bme280, riffelli2022wireless} on the Enviro+ board is a high-precision sensor that can measure temperature, pressure, and humidity. The placement of the sensor was deliberately made at the left side of the board with the intention of preventing any potential heat produced by the Raspberry Pi's CPU from reaching the sensor, which could interfere with the precision of the readings. The BME280 is commonly used for indoor environmental monitoring and can provide valuable information on the conditions of a home or other indoor space. There is a little slot right next to the sensor which can be useful to lessen the amount of heat that is generated from the Enviro+ board towards the direction of the sensor, further improving the accuracy of the readings.

\subsubsection{Light, and, proximity measurement}
The LTR-559 sensor \cite{riffelli2022wireless} on the Enviro and Enviro Plus sensor array can detect the amount of light in Lux. Lux represents the measure of the intensity of visible light. This sensor is useful for monitoring lighting conditions in indoor environments. Additionally, the LTR-559 sensor includes a proximity sensor, which can detect the presence of objects within a certain distance from the sensor. This feature can be used to create proximity-sensitive inputs, which can be useful in applications such as touchless control interfaces.

\begin{table}[ht]
\centering
\begin{tabular}{llllll}
\hline
\textbf{SENSOR\_TYPE} & \textbf{RPI\_ID} & \textbf{ROW\_COUNT} & \textbf{START\_DATE} & \textbf{END\_DATE} & \textbf{LOCATION} \\ \hline
Enviro   Plus & 20 & 283074 & 6/16/2021   18:01 & 1/17/2022   20:31 & House   B - Bedroom \\ \hline
Enviro   Plus & 21 & 309591 & 6/16/2021   18:42 & 3/11/2022   12:05 & House   B - Kitchen \\ \hline
Enviro & 22 & 279801 & 6/16/2021   19:07 & 1/12/2022   11:18 & House   C - Room B \\ \hline
Enviro & 23 & 280030 & 6/16/2021   19:26 & 1/12/2022   11:36 & House   C - Room A \\ \hline
Enviro   Plus & 30 & 438090 & 7/12/2021   21:56 & 7/1/2022   21:02 & House   C - Room A \\ \hline
Enviro & 37 & 85669 & 8/3/2021   3:26 & 7/6/2022   13:45 & House   A - Guest Room \\ \hline
Enviro   Plus & 39 & 242101 & 8/3/2021   2:28 & 7/6/2022   13:09 & House   A - Kitchen \\ \hline
Enviro & 41 & 43465 & 8/3/2021   3:44 & 9/7/2021   17:08 & House   A - Guest Room \\ \hline
Enviro & 45 & 352087 & 8/3/2021   3:59 & 7/6/2022   13:16 & House   A - Living Room \\ \hline
Enviro   Plus & 50 & 222585 & 8/3/2021   3:05 & 7/6/2022   12:56 & House   A - Master Bedroom \\ \hline
\end{tabular}
\caption{Sensor placements and record summary }
\label{tab:data_summary}
\end{table}

\subsubsection{Sound measurement}
The Enviro+ device features a miniature Micro-Electro-Mechanical System (MEMS) microphone \cite{gas_datasheet}, which facilitates the recording of audio and detection of noise levels \cite{loeppert2006sisonictm}. This functionality is particularly useful for monitoring and assessing the degree of noise pollution. The device is capable of detecting high, mid, and low sound levels, as well as measuring the amplitude of the sound.

\subsubsection{Gas measurement}
The MICS6814 \cite{de2020iot} is an analog gas sensor that is exclusively available in the Plus variant of the Enviro boards. This sensor can detect three distinct groups of gases categorized as reducing, oxidizing, and NH3 according to the datasheet. Notably, the MICS6814 can detect the presence of major gas or vapors like carbon monoxide (reducing), nitrogen dioxide (oxidizing), and ammonia (NH3), in addition to other gases including hydrogen, ethanol, and hydrocarbons.

\begin{figure}[ht]
\centering
\includegraphics[width=\textwidth]{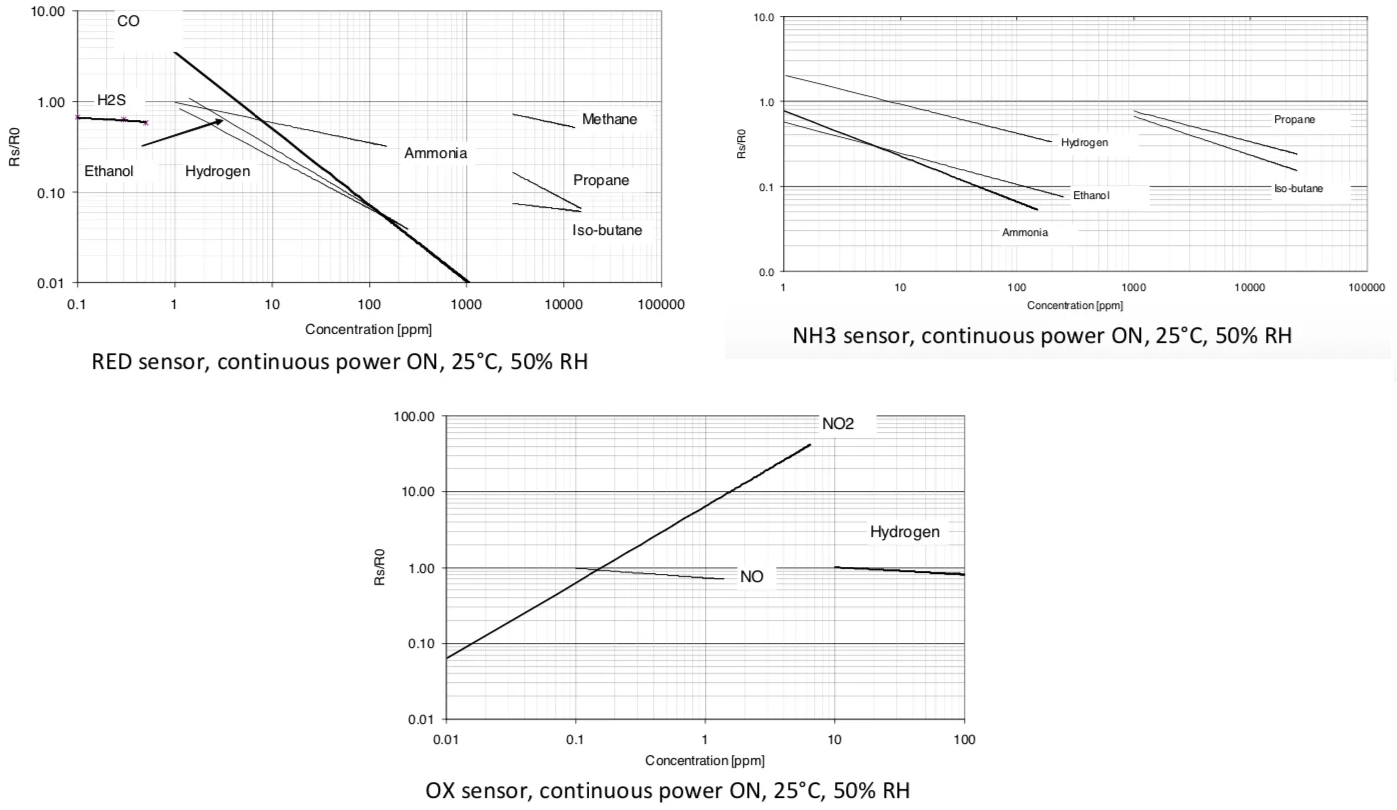}
\caption{MICS6814 reading chart for reducing, oxidising, and, ammonia \cite{gas_datasheet}}
\label{fig:gas_chart}
\end{figure}

As each gas group detected by the MICS6814 may consist of multiple gases, it is not possible to isolate or quantify any individual gas with precision. Therefore, the recommended approach for data interpretation is to record readings until they reach a steady state, establish a baseline, and then examine changes in relation to that baseline. This method provides a general indication of the air quality trend. Figure \ref{fig:gas_chart} presents a sample chart displaying MICS6814 readings for various gases. Further information on how to interpret these readings and the chart is available in the guide provided by Pimoroni \cite{gas_datasheet}. 

\subsection{Sensing Device Deployment}
The dataset presented in this study is part of a larger ongoing study primarily located in Virginia, USA. In this larger study, a total of $48$ sensing devices were deployed in different locations of $12$ individual households. Four sensing devices were placed in each house. The devices were placed with the acknowledged consent of the occupants living in the houses following the regulations of the Institutional Review Board (IRB), Virginia Tech. 
\\

The deployment of the sensing devices took place during the summer of 2021 and the devices have been collecting data since. However, not all the sensing devices are connected to the internet and such devices are collecting data in offline mode. The data from the offline devices can be retrieved manually with physical access to the devices once they are brought back. The current study presents data from $10$ devices of three individual houses. Some of these devices were retrieved from the households and some were online throughout the time span of a year. All data is currently present in the live server of the BDL system \cite{BDL_site}. The rest of the deployed sensing devices are still in the occupant houses and are still collecting data. 
\\

The dataset presented in this study was derived from a singular, non-random case study. The data collection process took place in three separate housing units, namely House A and House B, both located in Richmond, Virginia, and House C situated in Christiansburg, Virginia. Information regarding the placement locations of the sensing devices can be found in table \ref{tab:data_summary}. The sensors were installed in specific areas of each site, such as living rooms, kitchens, and bedrooms.
\\

\begin{figure}[ht]
\centering
\includegraphics[scale=0.40]{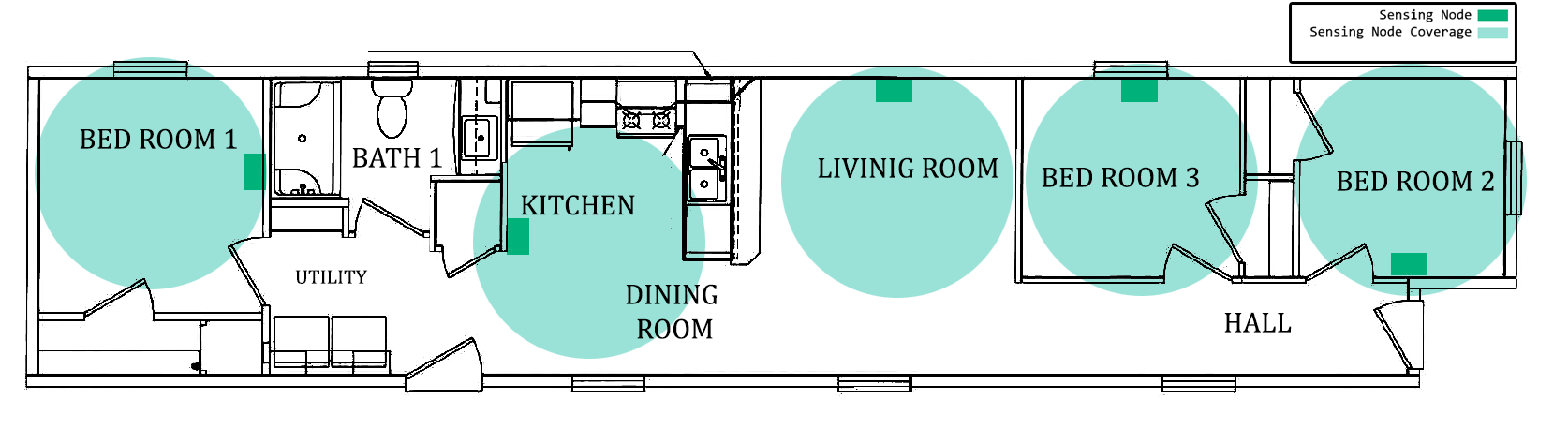}
\caption{Sensor deployment plan of House A}
\label{fig:plan_a}
\end{figure}

\begin{figure}[ht]
\centering
\includegraphics[scale=0.35]{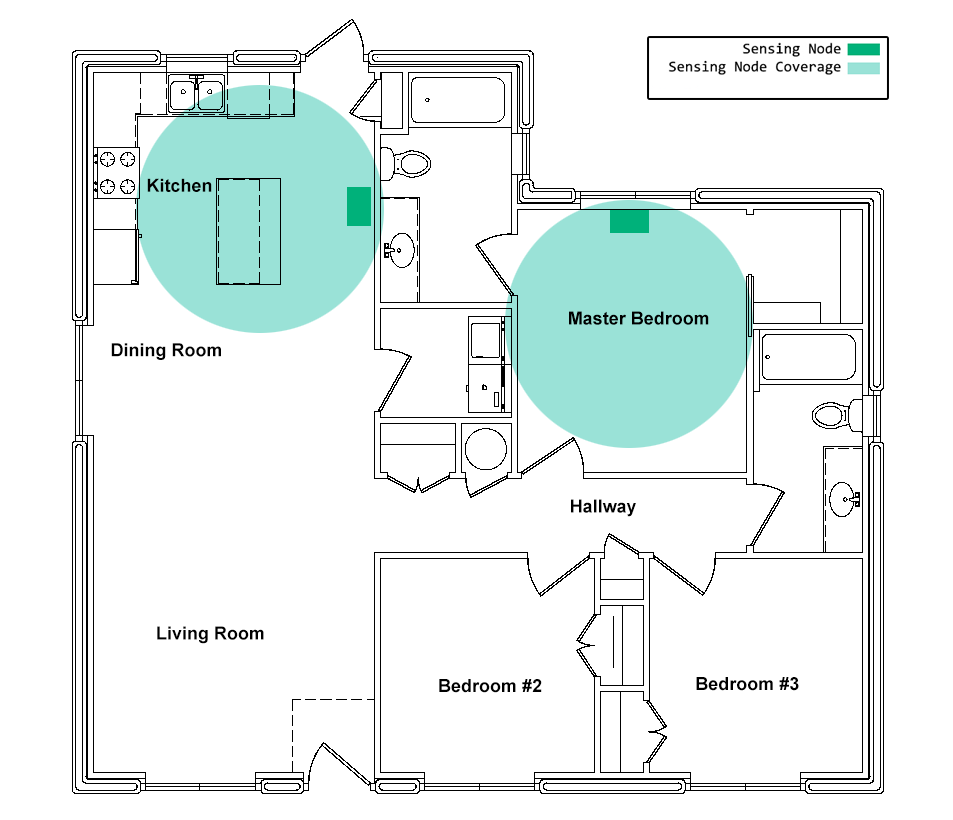}
\caption{Sensor deployment plan of House B}
\label{fig:plan_b}
\end{figure}

Figure \ref{fig:plan_a} depicts the layout of the sensing device deployment plan for House A. The green rectangles in the figure symbolize the sensing devices, while the light green circle that surrounds each sensing device represents the hypothetical coverage area of the device. These devices are directly connected to the central server through a Wi-Fi internet connection. Additionally, each device includes a local data backup in case of connectivity issues. These sensing devices are also portable and can be easily relocated to any location with access to power and network connection.

\subsection{Human data statement}
This research received approval from the Institutional Review Board of Virginia Tech (VT IRB 20-784 and 21-507), ensuring that the research coheres with ethical standards. Participants of this study were not subject to any known risks and informed consent was provided from all subjects.

%% file: contents/4_data_records.tex
\section{Data Records}
The dataset presented in this work along with the meta data has been made publicly available on the Open Source Framework repository \cite{Gao_Anik_2023}. It has also been made available on GitHub \cite{AnikGitHubBDL_Data1}. The collected dataset contains more than $2.5$ million records (2,536,493 exact) organized by the order of sensing devices. The dataset includes meta-data information about the collected data. This section describes the collected data along with the meta-data information. It also describes the data distribution and data organization.     

\subsection{Data description}
The data presented here has been collected through the Building Data Lite (BDL) sensing system \cite{anik2022cost, BDL_site}. The BDL site and all data can be publicly accessed. A total of 10 sensing nodes were deployed on various locations of 3 households. The deployment locations were randomly picked. Table \ref{tab:data_summary} provides detailed sensor placement information. It also includes information on the date range of the placement of the sensing devices along with the number of records each device collected during that period. Out of the 10 sensing nodes, 5 used the Enviro+ sensor array and the remaining 5 used the regular Enviro board.
\\

\begin{table}[ht]
\centering
\begin{tabular}{lllrrr}
\hline
\textbf{Attribute} & \textbf{Sensor} & \textbf{Unit} & \multicolumn{1}{l}{\textbf{House A (39)}} & \multicolumn{1}{l}{\textbf{House B (21)}} & \multicolumn{1}{l}{\textbf{House C (30)}} \\ \hline
proximity & LTR-559 & nm & 0 & 5 & 11 \\ \hline
humidity & BME280 & \%RH & 20.44 & 33.14 & 23.63 \\ \hline
pressure & BME280 & hPa & 1006.85 & 941.22 & 939.91 \\ \hline
light & LTR-559 & Lux & 4.22 & 2.33 & 0 \\ \hline
temperature & BME280 & °C & 29.01 & 26.55 & 26.59 \\ \hline
sound\_high & MEMS & dB & 30.43 & 30.02 & 30.03 \\ \hline
sound\_mid & MEMS & dB & 34.08 & 31.9 & 31.81 \\ \hline
sound\_low & MEMS & dB & 99.24 & 54.01 & 53.14 \\ \hline
sound\_amp & MEMS & dB & 27.52 & 20.28 & 20.12 \\ \hline
oxidised & MICS6814 & $k\Omega$ & 46.32 & 114.64 & 126.79 \\ \hline
reduced & MICS6814 & $k\Omega$ & 268.21 & 240.15 & 171.31 \\ \hline
nh3 & MICS6814 & $k\Omega$ & 78.2 & 89.63 & 99.16
\\ \hline
\end{tabular}
\caption{Reading units and median values of 3 devices in 3 houses. }
\label{tab:sensor_units}
\end{table}

Each record in this dataset either contains 12 attributes for Enviro boards or 15 for Enviro+ boards. The attributes are represented as columns in the CSV files. Columns 1 to 3 represent unique identification information and timestamps. Each of the rest of the columns represents an environmental attribute captured by the sensors. Columns 4 and 7 contain proximity and light data. The LTR-599 sensor is used to measure surrounding proximity and light level. Without making any physical contact, the sensor is capable of identifying the existence of objects that are close by. Proximity is recorded in the nanometer (nm) unit and light is measured in the Lux unit. Columns 5,6, and, 8 represent humidity, air pressure, and temperature data respectively. BME280 sensor is used to measure surrounding humidity, air pressure, and, temperature. These are measured in relative humidity (\%RH), Hectopascal (hPa), and, Degree Celcius (°C) respectively. 
\\

\begin{figure}[ht]
\centering
\includegraphics[width=\textwidth]{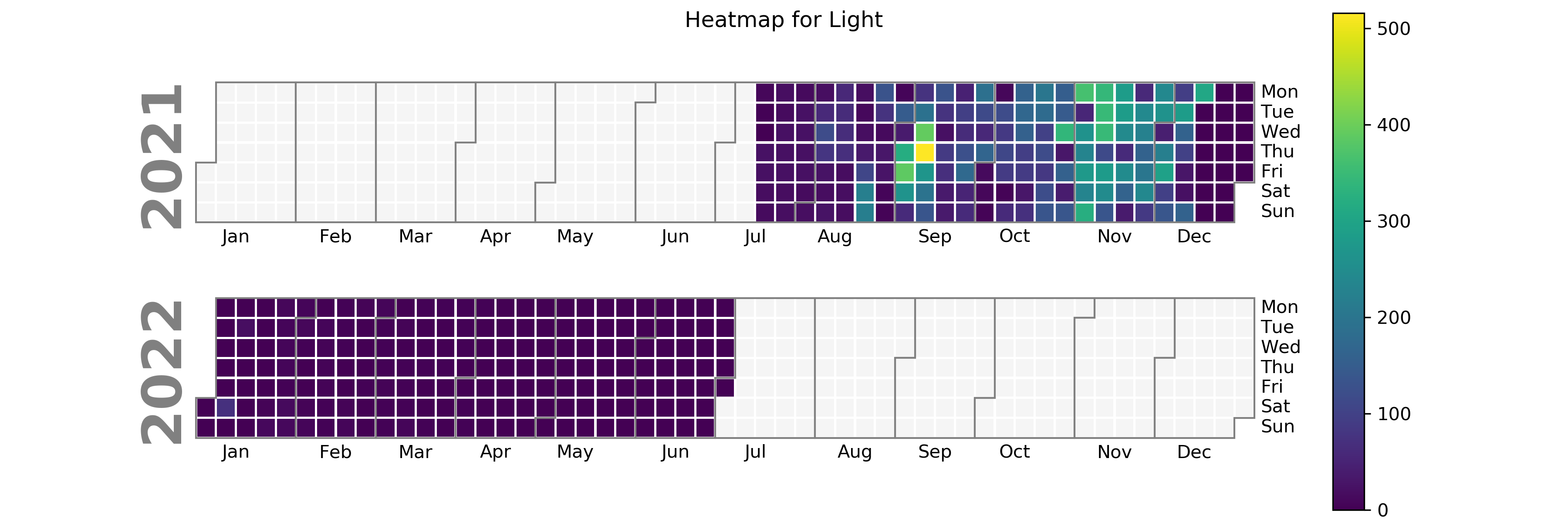}
\caption{Light heatmap data from Device \#30}
\label{fig:light_30}
\end{figure}

The Enviro boards feature a microelectromechanical systems (MEMS) microphone designed to capture sound events. The recorded sound data is represented in columns 9 through 11, which respectively correspond to high, mid, and low sound levels. Column 12 represents the amplitude of sound. The recorded sound data can be utilized for different purposes depending on the application. All sound levels are measured in decibels (dB) unit, which is the standard unit for measuring sound intensity.
\\

Columns 13 to 15 are only available in the Enviro+ CSV files. Columns 13, 14, and, 15 represent oxidised, reduced, and, ammonia (nh3) data collected from the MICS6814 sensor. The Enviro+ board includes an analog-to-digital converter (ADC) which is useful for interpreting gas sensor readings. The (ADC) takes the voltage readings generated by the sensor, and, converts them into resistances that can vary from several hundred Ohms to several tens of thousands of Ohms based on the different gas levels that are present. Gas particle levels are recorded in forms of resistance (Ohms). 
\\

Table \ref{tab:sensor_units} presents attribute details such as the sensors used to capture different environmental data and the units in which the data were recorded. Additionally, it shows the median values of each attribute from three devices located in three different households. The table reveals that the median temperature of the three houses falls within the range of 26°C to 29°C. Moreover, similarities in other attributes are also visible, such as the median sound amplitude ranging between 20 dB to 30 dB. House A (device 39) shows a median proximity reading of 0, indicating that during that reading, there was no movement around the sensing device resulting in zero proximity values. The low Lux level in the light recordings of all three houses indicates that the devices were possibly kept in shadows inside the houses. Although the median light reading of House C (device 30) is 0 Lux, the reading varies widely from 0 Lux to 500 Lux over the recording period of approximately one year. Figure \ref{fig:light_30} presents the light heatmap of device 30, confirming this variation.  
\\

\begin{figure}[ht]
\centering
\includegraphics[width=\textwidth]{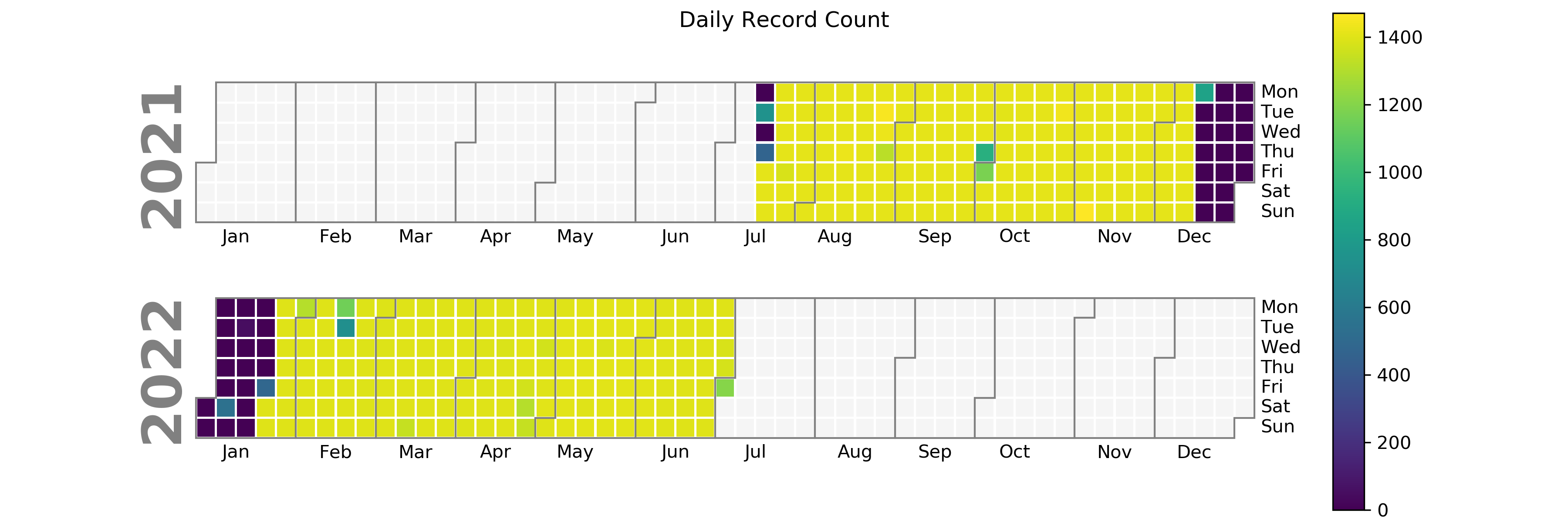}
\caption{Daily entry count heatmap of Device \#30 data}
\label{fig:count_30}
\end{figure}

The sensing devices used in this study were supposed to work 24\/7 throughout the entire duration. However, after analyzing the retrieved data, it came to light that, there were some time-frames when no data was collected. For example, figure \ref{fig:count_30} shows the heatmap of the count of daily entries from sensing device 30. Each device is expected to record $1440 (24 x 60)$ entries every day throughout the collection period except for the first and the last day. This is because the devices were programmed to capture one record every minute. In figure \ref{fig:count_30}, it can be seen that most of the days are bright yellow, representing the desired $1440$ entry count. However, there are a few days showing a lower entry count, somewhere around $1000$ marked in green. There is also a period from December 2021 to January 2022, where there are zero entries recorded. The sensing devices are programmed to continue data collection in the event of network failure or other connectivity issues. Thus the missing segments of figure \ref{fig:count_30} indicate that the sensing device was definitely turned off or cut out of power. Similar figures have been generated to provide a better understanding of the missing data and have been made available in the GitHub repository \cite{AnikGitHubBDL_Data1} along with the code to generate the images. 

\subsection{Meta-data}
The dataset includes additional meta-data files to provide supplementary information about the data collected by each sensing device. The meta-data are in text files named with the identification number of the corresponding sensing device. Each meta-data file contains 6 types of information as the following:

\begin{table}[ht]
\centering
\begin{tabular}{llllllllll}
\hline
\textbf{} & \textbf{proximity} & \textbf{humidity} & \textbf{pressure} & \textbf{light} & \textbf{oxidised} & \textbf{reduced} & \textbf{nh3} & \textbf{temperature} & \textbf{} \\ \hline
\textbf{mean} & 10.43935 & 22.52375 & 857.1613 & 59.54044 & 205.5385 & 183.4451 & 103.5953 & 26.49848 &  \\ \hline
\textbf{std} & 5.972486 & 6.419771 & 101.0621 & 87.02963 & 326.5619 & 36.8538 & 24.45422 & 2.037828 &  \\ \hline
\textbf{min} & 0 & 4.543291 & 649.5737 & 0 & 0.721915 & 7.505155 & 3.230769 & 11.40154 &  \\ \hline
\textbf{25\%} & 7 & 17.88105 & 742.3814 & 0 & 92.79227 & 154.2389 & 88.2623 & 25.1387 &  \\ \hline
\textbf{50\%} & 11 & 23.63057 & 939.9159 & 0 & 126.7893 & 171.3063 & 99.16373 & 26.59042 &  \\ \hline
\textbf{75\%} & 14 & 27.52975 & 947.0916 & 8.13295 & 222.733 & 212.9956 & 111.8474 & 27.89083 &  \\ \hline
\textbf{max} & 55 & 62.58616 & 961.5713 & 692.7939 & 3567.529 & 2877.333 & 1931.097 & 33.36806 &  \\ \hline
\end{tabular}
\caption{Sample data description section of a data file.}
\label{tab:data_description_sample}
\end{table}

\begin{itemize}
    \item ID: the ID section of the meta-data files represents the unique identification number assigned to the sensing device from which the data was collected. This number is recognized by the BDL central system and is specific to a single sensing device. 
    
    \item Data preview: this section of the meta-data file displays the first 5 rows of data collected by the sensing device, along with the names of each attribute. This provides a quick glimpse into the data recorded by the device.
    
    \item Columns : this section of the metadata file lists the names of the attributes recorded in the corresponding data file, separated by commas. For example, it may include attribute names such as 'id', 'date\_time', 'rpi\_id', 'proximity', 'humidity', 'pressure', 'light', 'oxidised', 'temperature', 'sound\_amp'. This section provides a quick reference for the types of data collected by the sensing device.
    
    \item Data info: the data info section provides information on the data structure of each attribute, including the index range and total number of entries. It also includes information on the data type and memory consumption of each attribute. This information can be useful for understanding the size and format of the data, as well as for optimizing memory usage and data processing.
    
    \item Data description: the data description section provides a statistical summary of the collected data for each attribute in a numeric manner. It includes the minimum, maximum, mean, median, and, standard deviation information for each attribute. Table \ref{tab:data_description_sample} shows an example of the data description section for one of the data files.
    
    \item Date range: this section in the meta-data file provides information on the time span of the data collected by the particular sensing device. Specifically, it shows the date and time of the first and last recorded data points. This information is important for understanding the temporal scope of the dataset and for identifying any potential gaps in the data collection. The date-time data is provided in the following format: "YYYY-MM-DD HH:MM:SS".
\end{itemize}

The authors have developed a Python script for generating the meta-data files from the CSV files downloaded from the BDL server. The script employs Pandas and CSV libraries for analyzing the data and generating the meta-data file in text format. The script has been made available on Github \cite{AnikGitHubBDL_Data1}.



%% file: contents/5_0_data_distribution.tex
\subsection{Data-distribution}
The data presented in this paper has been collected through the BDL system \cite{anik2022cost}. All data can be directly accessed through the live server \href{https://www.building-data-lite.com}{www.building-data-lite.com}. There is no need for any coding to access this data.  

\begin{figure}[ht]
\centering
\includegraphics[scale=0.45]{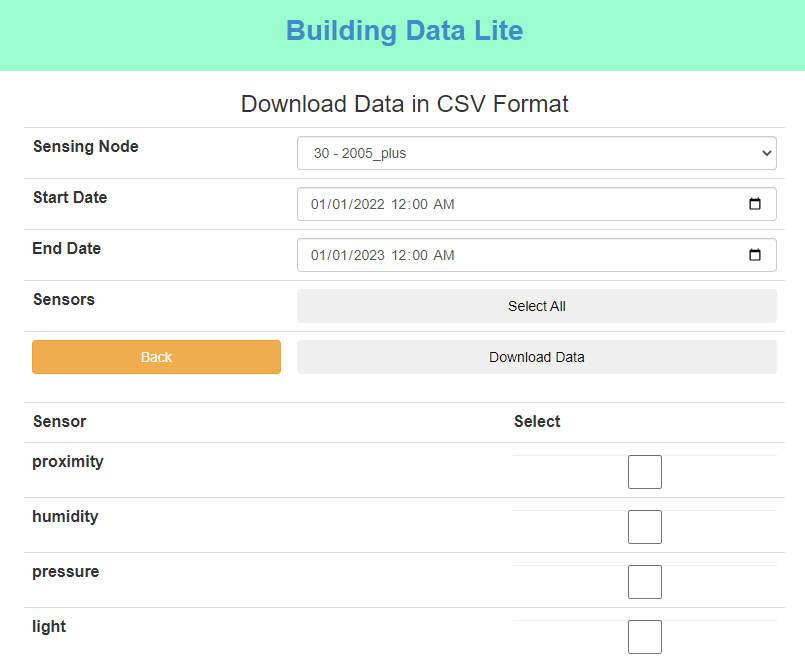}
\caption{Data download section of BDL System web interface}
\label{fig:data_download}
\end{figure}

Figure \ref{fig:data_download} shows a screenshot of the download page of the BDL website. A user can access this data through the guest account without any account credentials as all data currently available in the system is publicly accessible. However, the system requires user credentials to place sensing nodes in the system. The graphical user interface (GUI) of the BDL system accesses the central database through the central server of the BDL system. The system allows adding customized sensing devices through a code-free user interface. For the data collection process, the authors added 10 sensing nodes in the BDL system which have been collecting data since August 2021. Each sensing node is given a unique identification number by the BDL system during the installation time of the sensing device. As long as the devices are powered on, they can collect data without any interruptions. The devices are capable of continuing data collection even without any internet connection. In such circumstances, data is stored in local storage. Once the device gets re-connected to the internet, it synchronizes the local data with the live data available on the server. Afterward, it sends only the new data that was collected during the offline period back to the live server. The user can select the particular device ID through the download page to download the data collected by that device in CSV format. It also allows the user to download specific sensor data of the selected sensing device by selecting the checkboxes if needed. For example, the user can download only a single sensor data of a particular sensing device. 
\\

The dataset has been made publicly available on the Open Source Framework repository \cite{Gao_Anik_2023} and GitHub repository \cite{AnikGitHubBDL_Data1}. It includes a text guide on the data organization. The data is located in the data folder which includes two sub-folders named "plus" and "reg" respectively containing data from Enviro Plus and Enviro sensor arrays. The data files are in CSV format for ease of access. However, there is one file (ID 30), which was more than 100 MB in size. It is stored as a compressed file (only in GitHub \cite{AnikGitHubBDL_Data1}) and needs to be uncompressed prior to use. There is a folder named "meta\_data" which contains separate meta-data files of corresponding data files. Each meta-data file is named after the corresponding data file with a trailing "meta" keyword. Finally, there is a folder named "code" that contains two Python Notebook files. One is used for generating the meta-data files from the data file and another is used for validation of the collected data. 

%% file: contents/5_technical_validation.tex
\section{Technical Validation}


\begin{figure}[ht]
\centering
\includegraphics[scale=0.32]{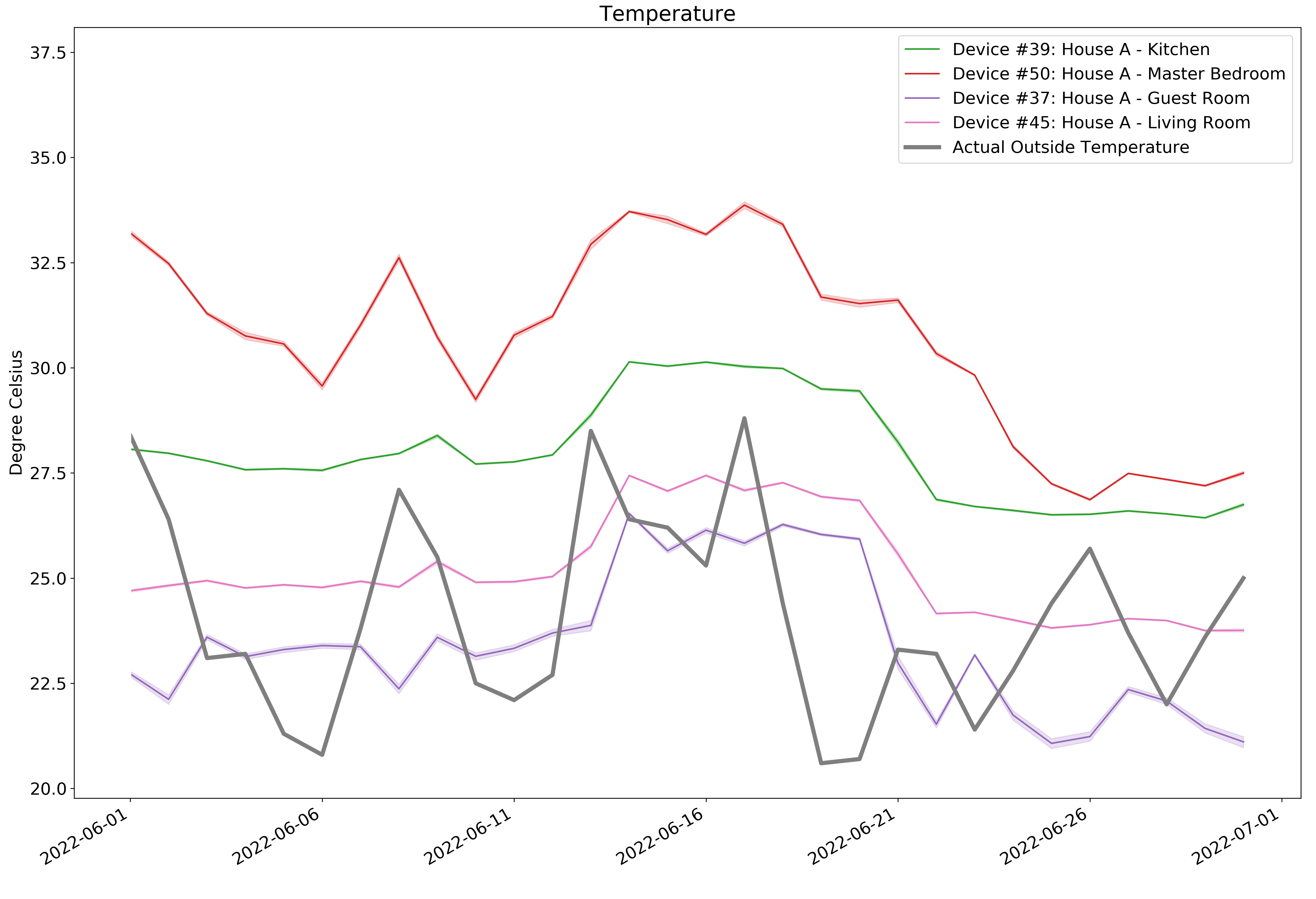}
\caption{Comparison of outside temperature data of June 2022 in Richmond, VA with indoor temperature records collected by the deployed sensors.}
\label{fig:temp_validation_june}
\end{figure}

\begin{figure}[ht]
\centering
\includegraphics[scale=0.32]{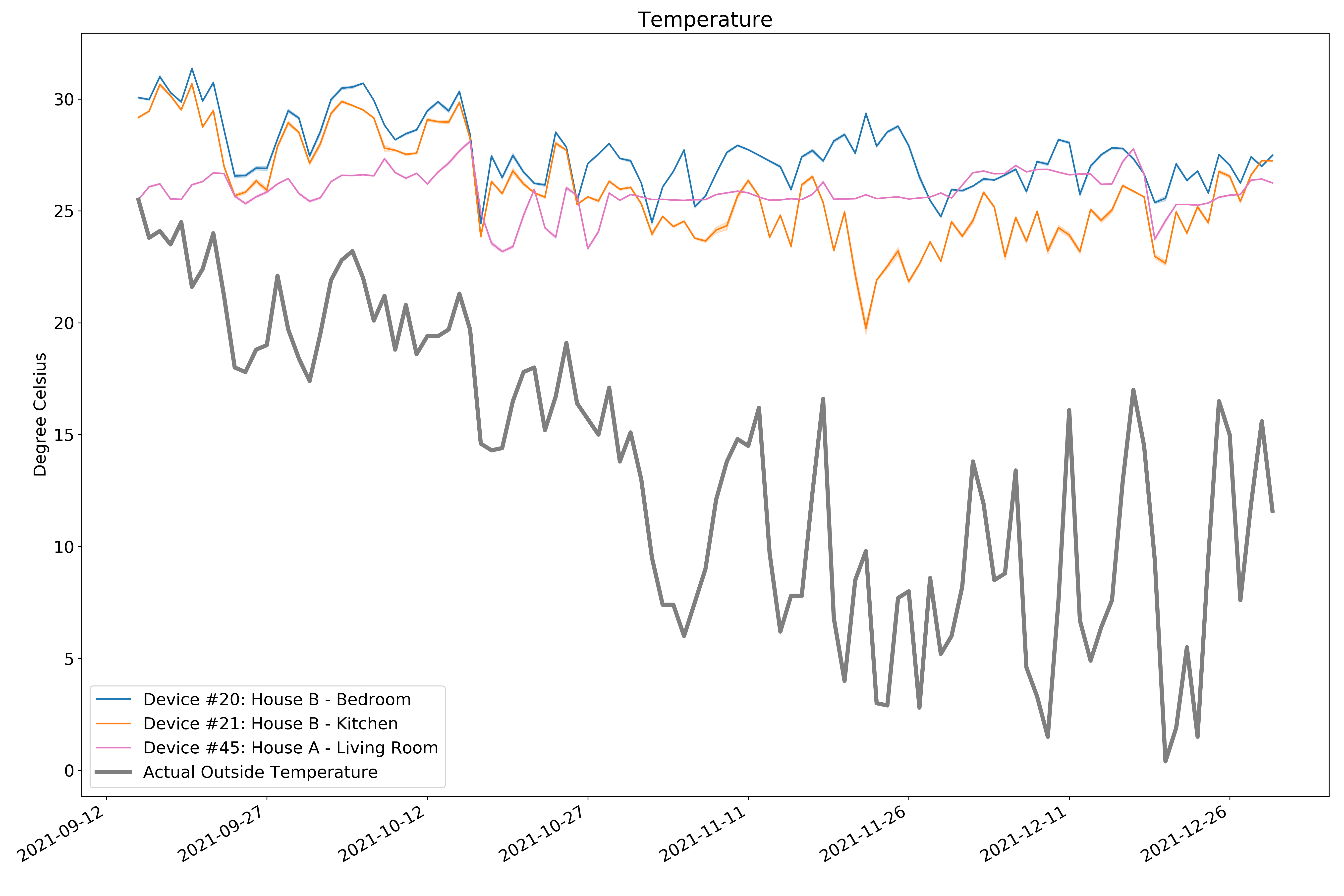}
\caption{Comparison of outside temperature data of September 2021 - December 2021 in Richmond, VA with indoor temperature records collected by the deployed sensors.}
\label{fig:temp_validation_long}
\end{figure}

\begin{figure}[ht]
\centering
\includegraphics[scale=0.32]{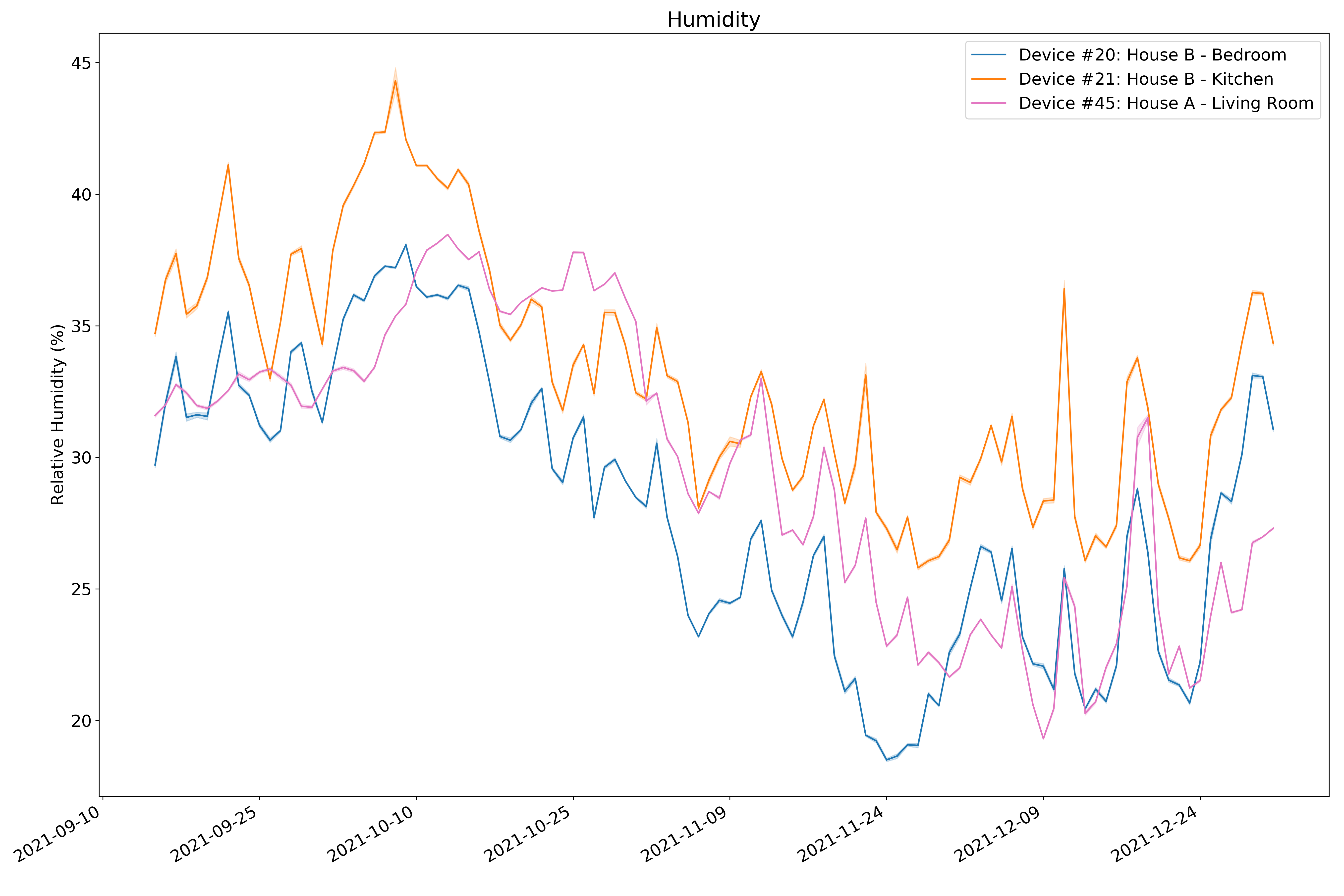}
\caption{Comparison of humidity records collected by the deployed sensors from September 2021 to December 2021 in Richmond, VA.}
\label{fig:humidity_validation_long}
\end{figure}

Ensuring the reliability of readings is of paramount importance, in cases where data is obtained through an individual case study. This is due to the lack of similar studies for comparison for the identification of any potential anomalies. The data presented in this description encompasses multiple domains of Building Performance Data (BPD) and requires the use of multiple instruments. However, due to the discrete nature of the experiment, it was not possible to directly verify every category of data presented in this dataset against real-world data. To maintain the integrity of the data, the authors frequently monitored the data being collected through the user interface of the BDL system \cite{BDL_site} during the recording period. In addition, the authors obtained actual weather data from Weather Underground \cite{weather_underground} and conducted a side-by-side comparison with the data collected by the BDL system to ensure its validity.


Figures \ref{fig:temp_validation_june} and \ref{fig:temp_validation_long} depict a comparison between the indoor temperature readings obtained from the placed sensing devices and the actual outside temperature records acquired from the Richmond International Airport Station through Weather Underground \cite{weather_underground}. Since capturing indoor temperature readings directly is not feasible, the authors used outside temperature as a reference scale. It is reasonable to expect a correlation between outdoor and indoor temperatures. The weather data used in this study was collected from the Weather Underground database \cite{weather_underground}.
\\

Figure \ref{fig:temp_validation_june} depicts the temperature recordings from June 2022 obtained from four sensing devices located in different areas of the house. The green line represents device 39 in the kitchen, the violet line shows device 37 in the guest room, the pink line represents device 45 in the living room, and the red line represents device 50 in the master bedroom. The bold black line indicates the actual outdoor temperature. The figures illustrate that the temperature readings of different devices demonstrate parallel patterns. Additionally, the data shows a striking resemblance between the actual outdoor temperature readings and the indoor temperature readings from different rooms. Moreover, temperature peaks observed in the outside temperature around June 10th, 12th, and 17th are also visible in the indoor sensor readings.
\\

Figure \ref{fig:temp_validation_long} presents a comparison between the actual outdoor temperature and indoor temperature readings from two distinct houses (House A and House B) situated in Richmond, VA. The graph displays data from three sensing devices, specifically device 20 located in the bedroom of House B (blue line), device 21 located in the kitchen of House B (orange line), and device 45 placed in the living room of House A (pink line). The outside temperature is indicated by a bold black line. The data in this figure pertains to a 4-month period, spanning from September 2021 to the end of December 2021. A noteworthy observation from this figure is that, despite the decline in the outdoor temperature during this period, the indoor temperature did not decrease significantly, presumably because of the use of room heaters. The indoor temperature lines also demonstrate a parallel nature, indicating a consistent pattern. Comparison of the collected data with the actual outdoor temperature displayed in figure \ref{fig:temp_validation_june} and figure \ref{fig:temp_validation_long} shows a plausible change pattern in temperature over different time periods, thus substantiating the reliability of the collected data.
\\

The validation of certain attributes, such as oxidised, reduced, or sound levels, is technically challenging due to the absence of a benchmark dataset for comparison. However, it is plausible for some attributes to have similar readings among the devices of the same house and also among devices of different houses located nearby. Humidity is such an attribute. Figure \ref{fig:humidity_validation_long} shows the humidity records collected by 3 sensing devices placed inside House A and House B. Device 20 is located in the bedroom of House B (blue line), device 21 is located in the kitchen of House B (orange line), and, device 45 is located in the living room of House A (pink line). Humidity records are recorded in the Relative Humidity (\% RH) unit. The indoor humidity is in general lower than the outside humidity. Here, a similarity in the humidity readings can be seen among different devices throughout the 4 months. Even devices placed in different houses demonstrate resemblance which indicates the validity of the collected data. 
\\

\begin{figure}[H]
\centering
\includegraphics[scale=0.32]{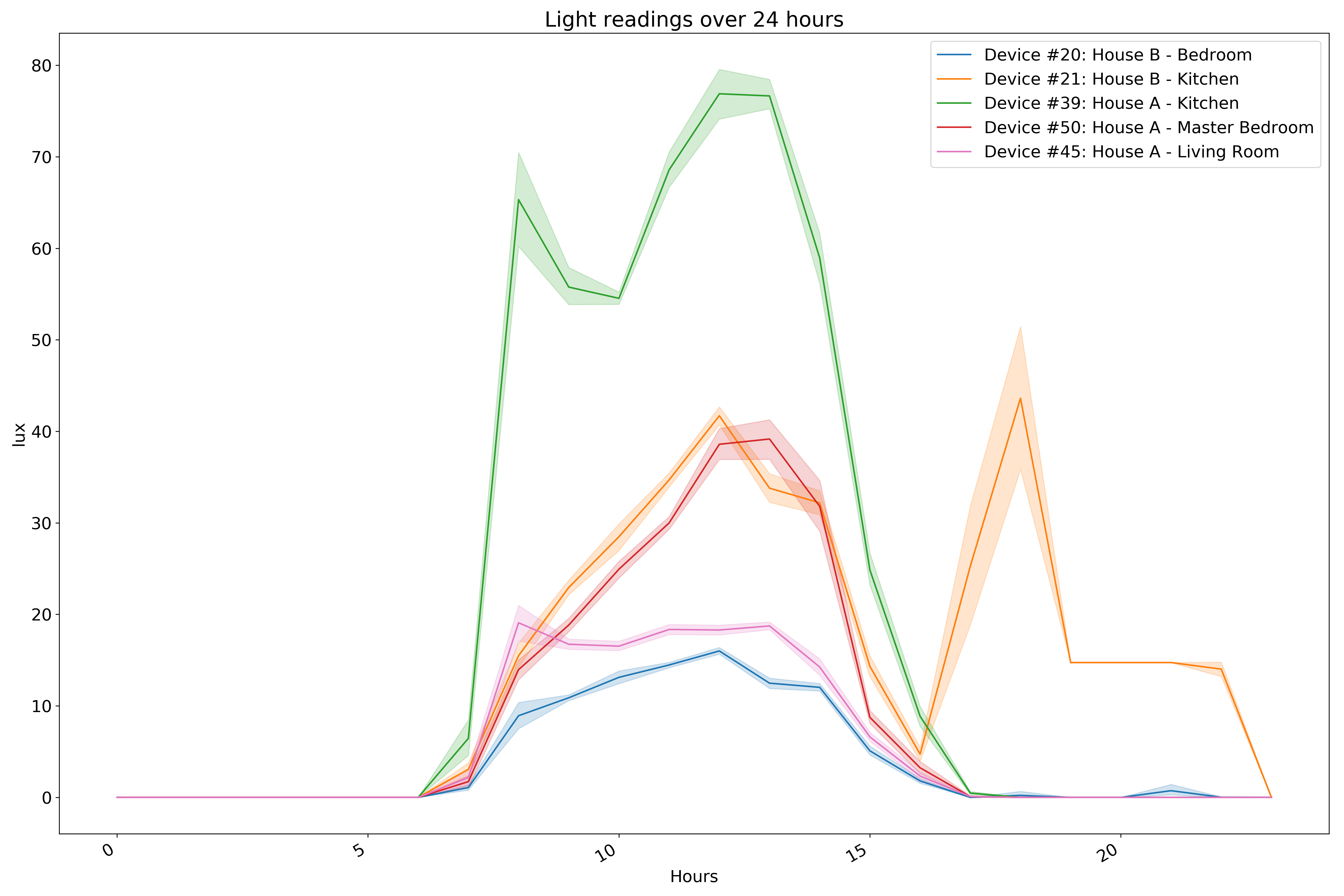}
\caption{Indoor light data of 24 hours of a day}
\label{fig:light_validation}
\end{figure}

Figure \ref{fig:light_validation} depicts the light measurements gathered by five sensing devices placed in different locations within House A and House B during a typical day (randomly selected). The blue and orange lines represent devices 20 and 21, located in the bedroom and kitchen of House B, respectively. Green, pink, and, red lines correspond to devices 39, 45, and 50, located in the kitchen, the living room, and, the master bedroom of House A, respectively. Light levels are measured in lux units. The figure indicates that during most of the night (12 AM to 6 AM), the sensors did not detect any light. The readings started increasing around 6:30 AM, reaching maximum values around midday (12 PM or 1 PM), and gradually decreased, reaching zero level around 7 PM. Notably, the orange line (kitchen of House B) shows light presence during the evening, indicating occupant activity after hours in the kitchen. The comparison of the readings among nearby devices further corroborates the validity of the data obtained through these sensing devices. While it is currently difficult to verify the remaining attributes, such as oxidized, reduced, or sound levels, as there is no benchmark dataset for comparison, future research works can be conducted to verify these attributes.

%% file: contents/6_usage_notes.tex
\section{Usage Notes}
The presented data has been collected through multiple similar sensing devices thus the data files include similar categories of data. The authors used the Pandas \cite{mckinney-proc-scipy-2010} and Matplotlib \cite{Hunter:2007} libraries of Python for data analysis and visualization. An example is available in the form of Jupyter Notebook \cite{Kluyver2016jupyter} on OSF \cite{Gao_Anik_2023} and GitHub \cite{AnikGitHubBDL_Data1}. However, other data analytical tools like MS Power PI or Tableau can also be used to analyze the data. 

%% file: contents/7_code_availability.tex
\section{Code availability}
The presented data was collected using the publicly available Building Data Lite (BDL) system \cite{anik2022cost} found at \href{https://www.building-data-lite.com}{www.building-data-lite.com} on 25 October 2022. No code was involved in generating the data and none is necessary to access or analyze it. To generate the metadata from the CSV data files, a single Python script was employed, which can be found at both OSF \cite{Gao_Anik_2023} and GitHub \cite{AnikGitHubBDL_Data1} repositories. This script utilizes the Pandas and CSV libraries of the Python system.

%% file: contents/8_author_contributions.tex
N.M. and X.G. supervised the project, provided insights on analyzing the collected data, and, guided the manuscript writing. X.G. installed the sensing devices in occupant housing. S.M.H.A. integrated the BDL system with the sensing devices, developed the scripts to validate the data, generate meta-data information, and, led the manuscript writing. All authors overlooked the data validation and reviewed the manuscript.

%% file: contents/9_competing_interests.tex
The authors declare no competing interests.

%% file: main.bbl
\begin{thebibliography}{10}
\urlstyle{rm}
\expandafter\ifx\csname url\endcsname\relax
  \def\url#1{\texttt{#1}}\fi
\expandafter\ifx\csname urlprefix\endcsname\relax\def\urlprefix{URL }\fi
\expandafter\ifx\csname doiprefix\endcsname\relax\def\doiprefix{DOI: }\fi
\providecommand{\bibinfo}[2]{#2}
\providecommand{\eprint}[2][]{\url{#2}}

\bibitem{hedge2014occupant}
\bibinfo{author}{Hedge, A.}, \bibinfo{author}{Miller, L.} \&
  \bibinfo{author}{Dorsey, J.}
\newblock \bibinfo{journal}{\bibinfo{title}{Occupant comfort and health in
  green and conventional university buildings}}.
\newblock {\emph{\JournalTitle{Work}}} \textbf{\bibinfo{volume}{49}},
  \bibinfo{pages}{363--372} (\bibinfo{year}{2014}).

\bibitem{mirzaei2020impact}
\bibinfo{author}{Mirzaei, N.}, \bibinfo{author}{Kamelnia, H.},
  \bibinfo{author}{Islami, S.~G.}, \bibinfo{author}{Kamyabi, S.} \&
  \bibinfo{author}{Assadi, S.~N.}
\newblock \bibinfo{journal}{\bibinfo{title}{The impact of indoor environmental
  quality of green buildings on occupants' health and satisfaction: A
  systematic review}}.
\newblock {\emph{\JournalTitle{Journal of Community Health Research}}}
  (\bibinfo{year}{2020}).

\bibitem{andargie2019review}
\bibinfo{author}{Andargie, M.~S.}, \bibinfo{author}{Touchie, M.} \&
  \bibinfo{author}{O'Brien, W.}
\newblock \bibinfo{journal}{\bibinfo{title}{A review of factors affecting
  occupant comfort in multi-unit residential buildings}}.
\newblock {\emph{\JournalTitle{Building and Environment}}}
  \textbf{\bibinfo{volume}{160}}, \bibinfo{pages}{106182}
  (\bibinfo{year}{2019}).

\bibitem{zhang2019healing}
\bibinfo{author}{Zhang, Y.}, \bibinfo{author}{Tzortzopoulos, P.} \&
  \bibinfo{author}{Kagioglou, M.}
\newblock \bibinfo{journal}{\bibinfo{title}{Healing built-environment effects
  on health outcomes: Environment--occupant--health framework}}.
\newblock {\emph{\JournalTitle{Building research \& information}}}
  \textbf{\bibinfo{volume}{47}}, \bibinfo{pages}{747--766}
  (\bibinfo{year}{2019}).

\bibitem{ghodrati2012green}
\bibinfo{author}{Ghodrati, N.}, \bibinfo{author}{Samari, M.} \&
  \bibinfo{author}{Shafiei, M. W.~M.}
\newblock \bibinfo{journal}{\bibinfo{title}{Green buildings impacts on
  occupants’ health and productivity}}.
\newblock {\emph{\JournalTitle{Journal of Applied Sciences Research}}}
  \textbf{\bibinfo{volume}{8}}, \bibinfo{pages}{4235--4241}
  (\bibinfo{year}{2012}).

\bibitem{mujan2019influence}
\bibinfo{author}{Mujan, I.}, \bibinfo{author}{An{\dj}elkovi{\'c}, A.~S.},
  \bibinfo{author}{Mun{\'c}an, V.}, \bibinfo{author}{Kljaji{\'c}, M.} \&
  \bibinfo{author}{Ru{\v{z}}i{\'c}, D.}
\newblock \bibinfo{journal}{\bibinfo{title}{Influence of indoor environmental
  quality on human health and productivity-a review}}.
\newblock {\emph{\JournalTitle{Journal of cleaner production}}}
  \textbf{\bibinfo{volume}{217}}, \bibinfo{pages}{646--657}
  (\bibinfo{year}{2019}).

\bibitem{li2014methods}
\bibinfo{author}{Li, Z.}, \bibinfo{author}{Han, Y.} \& \bibinfo{author}{Xu, P.}
\newblock \bibinfo{journal}{\bibinfo{title}{Methods for benchmarking building
  energy consumption against its past or intended performance: An overview}}.
\newblock {\emph{\JournalTitle{Applied Energy}}}
  \textbf{\bibinfo{volume}{124}}, \bibinfo{pages}{325--334}
  (\bibinfo{year}{2014}).

\bibitem{roth2020examining}
\bibinfo{author}{Roth, J.}, \bibinfo{author}{Lim, B.}, \bibinfo{author}{Jain,
  R.~K.} \& \bibinfo{author}{Grueneich, D.}
\newblock \bibinfo{journal}{\bibinfo{title}{Examining the feasibility of using
  open data to benchmark building energy usage in cities: A data science and
  policy perspective}}.
\newblock {\emph{\JournalTitle{Energy Policy}}} \textbf{\bibinfo{volume}{139}},
  \bibinfo{pages}{111327} (\bibinfo{year}{2020}).

\bibitem{quevedo2023applying}
\bibinfo{author}{Quevedo, T.}, \bibinfo{author}{Geraldi, M.} \&
  \bibinfo{author}{Melo, A.}
\newblock \bibinfo{journal}{\bibinfo{title}{Applying machine learning to
  develop energy benchmarking for university buildings in brazil}}.
\newblock {\emph{\JournalTitle{Journal of Building Engineering}}}
  \textbf{\bibinfo{volume}{63}}, \bibinfo{pages}{105468}
  (\bibinfo{year}{2023}).

\bibitem{robinson2017machine}
\bibinfo{author}{Robinson, C.} \emph{et~al.}
\newblock \bibinfo{journal}{\bibinfo{title}{Machine learning approaches for
  estimating commercial building energy consumption}}.
\newblock {\emph{\JournalTitle{Applied energy}}}
  \textbf{\bibinfo{volume}{208}}, \bibinfo{pages}{889--904}
  (\bibinfo{year}{2017}).

\bibitem{pipattanasomporn2020cu}
\bibinfo{author}{Pipattanasomporn, M.} \emph{et~al.}
\newblock \bibinfo{journal}{\bibinfo{title}{Cu-bems, smart building electricity
  consumption and indoor environmental sensor datasets}}.
\newblock {\emph{\JournalTitle{Scientific Data}}} \textbf{\bibinfo{volume}{7}},
  \bibinfo{pages}{241} (\bibinfo{year}{2020}).

\bibitem{tasgaonkar2022indoor}
\bibinfo{author}{Tasgaonkar, P.} \emph{et~al.}
\newblock \bibinfo{journal}{\bibinfo{title}{Indoor heat measurement data from
  low-income households in rural and urban south asia}}.
\newblock {\emph{\JournalTitle{Scientific Data}}} \textbf{\bibinfo{volume}{9}},
  \bibinfo{pages}{285} (\bibinfo{year}{2022}).

\bibitem{yoon2022datasets}
\bibinfo{author}{Yoon, Y.}, \bibinfo{author}{Jung, S.}, \bibinfo{author}{Im,
  P.} \& \bibinfo{author}{Gehl, A.}
\newblock \bibinfo{journal}{\bibinfo{title}{Datasets of a multizone office
  building under different hvac system operation scenarios}}.
\newblock {\emph{\JournalTitle{Scientific Data}}} \textbf{\bibinfo{volume}{9}},
  \bibinfo{pages}{775} (\bibinfo{year}{2022}).

\bibitem{gao2022understanding}
\bibinfo{author}{Gao, N.}, \bibinfo{author}{Marschall, M.},
  \bibinfo{author}{Burry, J.}, \bibinfo{author}{Watkins, S.} \&
  \bibinfo{author}{Salim, F.~D.}
\newblock \bibinfo{journal}{\bibinfo{title}{Understanding occupants’
  behaviour, engagement, emotion, and comfort indoors with heterogeneous
  sensors and wearables}}.
\newblock {\emph{\JournalTitle{Scientific Data}}} \textbf{\bibinfo{volume}{9}},
  \bibinfo{pages}{261} (\bibinfo{year}{2022}).

\bibitem{thorve2023high}
\bibinfo{author}{Thorve, S.} \emph{et~al.}
\newblock \bibinfo{journal}{\bibinfo{title}{High resolution synthetic
  residential energy use profiles for the united states}}.
\newblock {\emph{\JournalTitle{Scientific Data}}}
  \textbf{\bibinfo{volume}{10}}, \bibinfo{pages}{76} (\bibinfo{year}{2023}).

\bibitem{schwee2019room}
\bibinfo{author}{Schwee, J.~H.} \emph{et~al.}
\newblock \bibinfo{journal}{\bibinfo{title}{Room-level occupant counts and
  environmental quality from heterogeneous sensing modalities in a smart
  building}}.
\newblock {\emph{\JournalTitle{Scientific data}}} \textbf{\bibinfo{volume}{6}},
  \bibinfo{pages}{287} (\bibinfo{year}{2019}).

\bibitem{dong2022global}
\bibinfo{author}{Dong, B.} \emph{et~al.}
\newblock \bibinfo{journal}{\bibinfo{title}{A global building occupant behavior
  database}}.
\newblock {\emph{\JournalTitle{Scientific data}}} \textbf{\bibinfo{volume}{9}},
  \bibinfo{pages}{369} (\bibinfo{year}{2022}).

\bibitem{agee2021measured}
\bibinfo{author}{Agee, P.}, \bibinfo{author}{Nikdel, L.} \&
  \bibinfo{author}{Roberts, S.}
\newblock \bibinfo{journal}{\bibinfo{title}{A measured energy use, solar
  production, and building air leakage dataset for a zero energy commercial
  building}}.
\newblock {\emph{\JournalTitle{Scientific Data}}} \textbf{\bibinfo{volume}{8}},
  \bibinfo{pages}{299} (\bibinfo{year}{2021}).

\bibitem{paige2019fleece}
\bibinfo{author}{Paige, F.}, \bibinfo{author}{Agee, P.} \&
  \bibinfo{author}{Jazizadeh, F.}
\newblock \bibinfo{journal}{\bibinfo{title}{fleece, an energy use and occupant
  behavior dataset for net-zero energy affordable senior residential
  buildings}}.
\newblock {\emph{\JournalTitle{Scientific data}}} \textbf{\bibinfo{volume}{6}},
  \bibinfo{pages}{291} (\bibinfo{year}{2019}).

\bibitem{bashir2016towards}
\bibinfo{author}{Bashir, M.~R.} \& \bibinfo{author}{Gill, A.~Q.}
\newblock \bibinfo{title}{Towards an iot big data analytics framework: smart
  buildings systems}.
\newblock In \emph{\bibinfo{booktitle}{2016 IEEE 18th International Conference
  on High Performance Computing and Communications; IEEE 14th International
  Conference on Smart City; IEEE 2nd International Conference on Data Science
  and Systems (HPCC/SmartCity/DSS)}}, \bibinfo{pages}{1325--1332}
  (\bibinfo{organization}{IEEE}, \bibinfo{year}{2016}).

\bibitem{baghalzadeh2022internet}
\bibinfo{author}{Baghalzadeh~Shishehgarkhaneh, M.}, \bibinfo{author}{Keivani,
  A.}, \bibinfo{author}{Moehler, R.~C.}, \bibinfo{author}{Jelodari, N.} \&
  \bibinfo{author}{Roshdi~Laleh, S.}
\newblock \bibinfo{journal}{\bibinfo{title}{Internet of things (iot), building
  information modeling (bim), and digital twin (dt) in construction industry: A
  review, bibliometric, and network analysis}}.
\newblock {\emph{\JournalTitle{Buildings}}} \textbf{\bibinfo{volume}{12}},
  \bibinfo{pages}{1503} (\bibinfo{year}{2022}).

\bibitem{tang2019review}
\bibinfo{author}{Tang, S.}, \bibinfo{author}{Shelden, D.~R.},
  \bibinfo{author}{Eastman, C.~M.}, \bibinfo{author}{Pishdad-Bozorgi, P.} \&
  \bibinfo{author}{Gao, X.}
\newblock \bibinfo{journal}{\bibinfo{title}{A review of building information
  modeling (bim) and the internet of things (iot) devices integration: Present
  status and future trends}}.
\newblock {\emph{\JournalTitle{Automation in Construction}}}
  \textbf{\bibinfo{volume}{101}}, \bibinfo{pages}{127--139}
  (\bibinfo{year}{2019}).

\bibitem{zakaria2018wireless}
\bibinfo{author}{Zakaria, N.~A.} \emph{et~al.}
\newblock \bibinfo{journal}{\bibinfo{title}{Wireless internet of things-based
  air quality device for smart pollution monitoring}}.
\newblock {\emph{\JournalTitle{Int. J. Adv. Comput. Sci. Appl}}}
  \textbf{\bibinfo{volume}{9}}, \bibinfo{pages}{65--69} (\bibinfo{year}{2018}).

\bibitem{marques2019cost}
\bibinfo{author}{Marques, G.} \& \bibinfo{author}{Pitarma, R.}
\newblock \bibinfo{journal}{\bibinfo{title}{A cost-effective air quality
  supervision solution for enhanced living environments through the internet of
  things}}.
\newblock {\emph{\JournalTitle{Electronics}}} \textbf{\bibinfo{volume}{8}},
  \bibinfo{pages}{170} (\bibinfo{year}{2019}).

\bibitem{anik2022cost}
\bibinfo{author}{Anik, S. M.~H.}, \bibinfo{author}{Gao, X.},
  \bibinfo{author}{Meng, N.}, \bibinfo{author}{Agee, P.~R.} \&
  \bibinfo{author}{McCoy, A.~P.}
\newblock \bibinfo{journal}{\bibinfo{title}{A cost-effective, scalable, and
  portable iot data infrastructure for indoor environment sensing}}.
\newblock {\emph{\JournalTitle{Journal of Building Engineering}}}
  \bibinfo{pages}{104027} (\bibinfo{year}{2022}).

\bibitem{AnikGitHub2021}
\bibinfo{author}{Anik, S. M.~H.}
\newblock \emph{\bibinfo{title}{BDL Project Repository}}
  (\bibinfo{year}{accessed October 1, 2022}).
\newblock \bibinfo{note}{Url:
  \href{https://github.com/anik801/data_collection}{https://github.com/anik801/data\_collection}}.

\bibitem{MariaDBFoundation2016}
\bibinfo{author}{{MariaDB Foundation}}.
\newblock \bibinfo{title}{{About Mariadb}} (\bibinfo{year}{2016}).

\bibitem{kenler2015mariadb}
\bibinfo{author}{Kenler, E.} \& \bibinfo{author}{Razzoli, F.}
\newblock \emph{\bibinfo{title}{MariaDB Essentials}} (\bibinfo{publisher}{Packt
  Publishing Ltd}, \bibinfo{year}{2015}).

\bibitem{bartholomew2012mariadb}
\bibinfo{author}{Bartholomew, D.}
\newblock \bibinfo{journal}{\bibinfo{title}{Mariadb vs. mysql}}.
\newblock {\emph{\JournalTitle{Dostopano}}} \textbf{\bibinfo{volume}{7}},
  \bibinfo{pages}{2014} (\bibinfo{year}{2012}).

\bibitem{gas_datasheet}
\bibinfo{author}{Macdonald, S.}
\newblock \bibinfo{title}{Getting started with enviro+} (\bibinfo{year}{2019}).

\bibitem{sensortec2015bme280}
\bibinfo{author}{Sensortec, B.}
\newblock \bibinfo{journal}{\bibinfo{title}{Bme280 combined humidity and
  pressure sensor}}.
\newblock {\emph{\JournalTitle{Bosch Sensortec}}}  (\bibinfo{year}{2015}).

\bibitem{riffelli2022wireless}
\bibinfo{author}{Riffelli, S.}
\newblock \bibinfo{journal}{\bibinfo{title}{A wireless indoor environmental
  quality logger processing the indoor global comfort index}}.
\newblock {\emph{\JournalTitle{Sensors}}} \textbf{\bibinfo{volume}{22}},
  \bibinfo{pages}{2558} (\bibinfo{year}{2022}).

\bibitem{loeppert2006sisonictm}
\bibinfo{author}{Loeppert, P.~V.} \& \bibinfo{author}{Lee, S.~B.}
\newblock \bibinfo{title}{Sisonictm-the first commercialized mems microphone}.
\newblock In \emph{\bibinfo{booktitle}{Proceedings of the Solid-State Sensors,
  Actuators, and Microsystems Workshop}}, \bibinfo{pages}{27--30}
  (\bibinfo{year}{2006}).

\bibitem{de2020iot}
\bibinfo{author}{de~Medeiros, H. P.~L.} \& \bibinfo{author}{Gir{\~a}o, G.}
\newblock \bibinfo{title}{An iot-based air quality monitoring platform}.
\newblock In \emph{\bibinfo{booktitle}{2020 IEEE international smart cities
  conference (ISC2)}}, \bibinfo{pages}{1--6} (\bibinfo{organization}{IEEE},
  \bibinfo{year}{2020}).

\bibitem{BDL_site}
\bibinfo{author}{Anik, S. M.~H.}
\newblock \emph{\bibinfo{title}{Building Data Lite}} (\bibinfo{year}{accessed
  March 1, 2023}).
\newblock \bibinfo{note}{Url:
  \href{https://www.building-data-lite.com}{https://www.building-data-lite.com}}.

\bibitem{Gao_Anik_2023}
\bibinfo{author}{Gao, X.} \& \bibinfo{author}{Anik, M.~H.}
\newblock \bibinfo{title}{A comprehensive indoor environment dataset from
  single-family houses in the us}, \url{https://doi.org/10.17605/OSF.IO/BAEW7}
  (\bibinfo{year}{2023}).

\bibitem{AnikGitHubBDL_Data1}
\bibinfo{author}{Anik, S. M.~H.}
\newblock \emph{\bibinfo{title}{BDL Collected Data Phase 1}}
  (\bibinfo{year}{accessed March 1, 2023}).
\newblock \bibinfo{note}{Url:
  \href{https://github.com/anik801/BDL_data_1}{https://github.com/anik801/BDL\_data\_1}}.

\bibitem{weather_underground}
\bibinfo{author}{Underground, W.}
\newblock \bibinfo{title}{Henrico, va weather history} (\bibinfo{year}{2021}).

\bibitem{mckinney-proc-scipy-2010}
\bibinfo{author}{{W}es {M}c{K}inney}.
\newblock \bibinfo{title}{{D}ata {S}tructures for {S}tatistical {C}omputing in
  {P}ython}.
\newblock In \bibinfo{editor}{{S}t\'efan van~der {W}alt} \&
  \bibinfo{editor}{{J}arrod {M}illman} (eds.)
  \emph{\bibinfo{booktitle}{{P}roceedings of the 9th {P}ython in {S}cience
  {C}onference}}, \bibinfo{pages}{56 -- 61}, \url{10.25080/Majora-92bf1922-00a}
  (\bibinfo{year}{2010}).

\bibitem{Hunter:2007}
\bibinfo{author}{Hunter, J.~D.}
\newblock \bibinfo{journal}{\bibinfo{title}{Matplotlib: A 2d graphics
  environment}}.
\newblock {\emph{\JournalTitle{Computing in Science \& Engineering}}}
  \textbf{\bibinfo{volume}{9}}, \bibinfo{pages}{90--95},
  \url{10.1109/MCSE.2007.55} (\bibinfo{year}{2007}).

\bibitem{Kluyver2016jupyter}
\bibinfo{author}{Kluyver, T.} \emph{et~al.}
\newblock \bibinfo{title}{Jupyter notebooks -- a publishing format for
  reproducible computational workflows}.
\newblock In \bibinfo{editor}{Loizides, F.} \& \bibinfo{editor}{Schmidt, B.}
  (eds.) \emph{\bibinfo{booktitle}{Positioning and Power in Academic
  Publishing: Players, Agents and Agendas}}, \bibinfo{pages}{87 -- 90}
  (\bibinfo{organization}{IOS Press}, \bibinfo{year}{2016}).

\end{thebibliography}
